\documentclass[preprint,amsmath,amssymb,11pt,english,aps,showpacs,showkeys]{revtex4}
\usepackage{graphicx}
\usepackage{graphics}
\usepackage{amsmath}
\usepackage{dcolumn}
\usepackage{amssymb}
\usepackage{bm}
\usepackage[latin1]{inputenc}

\begin{document}
\title{On the interaction of the Dirac oscillator with the Aharonov-Casher system in topological defect backgrounds}
\author{K. Bakke}
\email{kbakke@fisica.ufpb.br}
\affiliation{Departamento de F\'isica, Universidade Federal da Para\'iba, Caixa Postal 5008, 58051-970, Jo\~ao Pessoa, PB, Brazil.}

\author{C. Furtado}
\email{furtado@fisica.ufpb.br} 
\affiliation{Departamento de F\'{\i}sica, Universidade Federal da Para\'{\i}ba, Caixa Postal 5008, 58051-970, Jo\~{a}o Pessoa, PB, Brazil}
 
\begin{abstract}
In this paper, we study the influence of the Aharonov-Casher effect [Y. Aharonov and A. Casher, Phys. Rev. Lett. {\bf 53}, 319 (1984).] on the Dirac oscillator in three different scenarios of general relativity: the Minkowski spacetime, the cosmic string spacetime and the cosmic dislocation spacetime. In this way, we solve the Dirac equation and obtain the energy levels for bound states and the Dirac spinors for positive-energy solutions. We show that the relativistic energy levels depend on the Aharonov-Casher geometric phase. We also discuss the influence of curvature and torsion on the relativistic energy levels and the Dirac spinors due to the topology of the cosmic string and cosmic dislocation spacetimes.  
\end{abstract}


\maketitle

\section{Introduction}

The significance of the electromagnetic potential in quantum mechanics demonstrated by Aharonov and Bohm \cite{ab} has opened new discussions and discovered interesting quantum effects \cite{dab,pesk,zei2,alman} associated with geometric phases. At present days, geometric phases have been studied in condensed matter physics \cite{book}, holonomic quantum computation \cite{vedral,bf20} and persistent currents in mesoscopic rings and quantum dots \cite{by,ring1,ring2,bf18}. An interesting study of geometric phases was proposed by Aharonov and Casher \cite{ac} for a neutral particle, which is known as the Aharonov-Casher effect. The Aharonov-Casher effect \cite{ac} corresponds to the appearance of a geometric quantum phase in the wave function of a neutral particle with a permanent magnetic dipole moment interacting with a radial electric field produced by a uniform linear distribution of electric charges, that is,
\begin{eqnarray}
\phi_{\mathrm{AC}}=\frac{1}{\hbar c}\oint\left(\vec{\mu}\times\vec{E}\right)_{\nu}\,dx^{\nu}=\pm2\pi\,\frac{\mu\lambda}{\hbar c}.
\label{1}
\end{eqnarray} 

The study of geometric quantum phases for permanent dipole moments has attracted a great deal of attention in recent decades \cite{anan,sil}. The dual effect of the Aharonov-Casher effect corresponds to the arising of a geometric phase from the interaction between a permanent electric dipole moment and a radial magnetic field which is called the He-McKellar-Wilkens effect \cite{hmw}. Analogous effects to the Aharonov-Casher effect have been obtained in the noncommutative quantum mechanics \cite{fur2}, in the Lorentz-symmetry violation background \cite{belich}, in the presence of topological defects \cite{bf1,bf2} and in noninertial reference frames \cite{bf4}.

Another interesting relativistic quantum system that has attracted a great deal of attention recently is the Dirac oscillator \cite{osc1}. Recent works have shown the impossibility of recovering the harmonic oscillator Hamiltonian from the nonrelativistic limit of the Dirac equation that describes the interaction between a relativistic spin-half particle with the harmonic oscillator potential. This impossibility occurs due to the presence of a quadratic potential \cite{osc1,do2,osc2} in the second order differential equation. Therefore, in order to solve this problem, the relativistic harmonic oscillator has been investigated recently by introducing scalar and vector potentials which are quadratics in coordinates \cite{prc} and by introducing a new coupling into the Dirac equation \cite{osc1}. In particular, the new coupling proposed in Ref. \cite{osc1} is introduced in such a way that the Dirac equation remains linear in both spatial coordinates and momenta, and recover the Schr\"odinger equation for a harmonic oscillator in the nonrelativistic limit of the Dirac equation. This new coupling introduced into the Dirac equation corresponds to the Dirac oscillator \cite{osc1}, which is given by 
\begin{eqnarray}
\vec{p}\rightarrow\vec{p}-im\omega\rho\,\hat{\beta}\,\hat{\rho},
\label{2}
\end{eqnarray}
where $m$ is the mass of the Dirac neutral particle, $\omega$ is the oscillator frequency, $\hat{\beta}$ is one of the standard Dirac matrices, and $\hat{\rho}$ is a unit vector in the radial direction. 

Recently, an analogy between the Dirac oscillator and the Jaynes-Cummings model \cite{jay} has been made in Refs. \cite{jay2,osc3} and extended to studies of the Ramsey-interferometry effect \cite{osc6}. In Ref. \cite{jay2}, Bermudez {\it et al} obtained an exact mapping of the Dirac oscillator onto the Jaynes-Cummings model describing the interaction of a two-level atom with a quantized single-mode field. The authors of Ref. \cite{osc3} investigated, in the limit the strong spin-orbit coupling of the Dirac oscillator, the entanglement of the spin with the orbital motion is produced in a way similar to what is observed in the model of the Jaynes-Cummings model. The Dirac oscillator has also been analyzed in a series of physical systems, such as in the presence of an external magnetic field \cite{osc9}, in studies of covariance properties \cite{moreno}, in the point of view of the Lie algebra \cite{quesnemo}, in a thermal bath \cite{osc10}, in the hidden supersymmetry \cite{benitez,moreno,quesne2}, by using the shape invariant method \cite{quesne}, conformal invariance properties \cite{romerom}, in the presence of external magnetic fields, under the influence of noninertial effects \cite{b10} and in the presence of the Aharonov-Bohm quantum flux \cite{ferk,luis,rojas,victor}. Furthermore, the Dirac oscillator was investigated by one of us for a system of a charged particle interacting with a topological defect \cite{josevi}.

The purpose of this paper is to study the influence of the Aharonov-Casher effect \cite{ac} on the Dirac oscillator \cite{osc1} in three different scenarios of general relativity: the Minkowski spacetime, the cosmic string spacetime and the cosmic dislocation spacetime. Hence, we solve the Dirac equation for the Dirac oscillator under the influence of the Aharonov-Casher effect \cite{ac} by showing the dependence of the relativistic energy levels on the Aharonov-Casher geometric phase. We also discuss the influence of curvature and torsion on the relativistic energy levels due to the topology of the cosmic string spacetime and the cosmic dislocation spacetime. Finally, for each case, we obtain the Dirac spinors for positive-energy solutions and discuss the arising of persistent spin currents \cite{by,ring1,ring2,bf18}.

The structure of this paper is: in section II, we study the the influence of the Aharonov-Casher effect \cite{ac} on the Dirac oscillator \cite{osc1} in the Minkowski spacetime; in section III, we discuss the influence of curvature due to the topology of the cosmic string spacetime on the Dirac oscillator interacting with the Aharonov-Casher system; in section IV, we discuss the the influence of torsion due to the topology of the cosmic dislocation spacetime on the Dirac oscillator interacting with the Aharonov-Casher system; in section V, we discuss some applications by obtaining the persistent spin currents; in section VI, we present our conclusions.

\section{the Dirac oscillator and the relativistic Aharonov-Casher system}

In this section, we study the influence of the Aharonov-Casher effect \cite{ac} on the Dirac oscillator \cite{osc1} in the Minkowski spacetime. We begin by introducing the relativistic quantum dynamics of a neutral particle possessing a permanent magnetic dipole moment interacting with electric and magnetic fields in the Minkowski spacetime. In the following, we introduce the Dirac oscillator coupling into the Dirac equation and solve the Dirac equation. 

The relativistic quantum dynamics of a neutral particle which describes the relativistic Aharonov-Casher system \cite{ac} is given by the introduction of a nonminimal coupling into the Dirac equation \cite{anan} given by (in Cartesian coordinates)
\begin{eqnarray}
i\gamma^{\mu}\,\partial_{\mu}\rightarrow i\gamma^{\mu}\,\partial_{\mu}+\frac{\mu}{2}\Sigma^{\mu\nu}\,F_{\mu\nu}\left(x\right),
\label{2.1}
\end{eqnarray}
where we consider the units $\hbar=c=1$ from now on. We have in (\ref{2.1}) that $\mu$ corresponds to the magnetic dipole moment of the neutral particle, $F_{\mu\nu}\left(x\right)$ corresponds to the electromagnetic tensor, whose components are defined as $F_{0i}=-F_{i0}=E_{i}$ and $F_{ij}=-F_{ji}=-\epsilon_{ijk}\,B^{k}$, and $\Sigma^{ab}=\frac{i}{2}\left[\gamma^{a},\gamma^{b}\right]$. The $\gamma^{a}$ matrices correspond to the Dirac matrices in the Minkowski spacetime \cite{greiner}:
\begin{eqnarray}
\gamma^{0}=\hat{\beta}=\left(
\begin{array}{cc}
1 & 0 \\
0 & -1 \\
\end{array}\right);\,\,\,\,\,\,
\gamma^{i}=\hat{\beta}\,\hat{\alpha}^{i}=\left(
\begin{array}{cc}
 0 & \sigma^{i} \\
-\sigma^{i} & 0 \\
\end{array}\right);\,\,\,\,\,\,\Sigma^{i}=\left(
\begin{array}{cc}
\sigma^{i} & 0 \\
0 & \sigma^{i} \\	
\end{array}\right),
\label{2.2}
\end{eqnarray}
with $\gamma^{a}\gamma^{b}+\gamma^{b}\gamma^{a}=-2\eta^{ab}$, $\vec{\Sigma}$ being the spin vector and $\sigma^{i}$ being the Pauli matrices. The tensor $\eta^{ab}=\mathrm{diag}(- + + +)$ is the Minkowski tensor.

Now, let us introduce the Aharonov-Casher effect \cite{ac} and the cylindrical symmetry of this system. The Aharonov-Casher effect \cite{ab} corresponds to the appearance of a geometric quantum phase in the wave function of a neutral particle given by the interaction between the magnetic dipole moment of the neutral particle and a radial electric field produced by a linear distribution of electric charges perpendicular to the plane of motion of the neutral particle, that is, $\vec{E}=\frac{\lambda}{\rho}\,\hat{\rho}$ (where $\hat{\rho}$ is a unit vector on the radial direction, $\rho^{2}=x^{2}+y^{2}$ and $\lambda$ is a linear electric charge distribution along the $z$ axis). Moreover, by introducing the coupling that describes the Dirac oscillator $\vec{p}\rightarrow\vec{p}-im\omega\rho\,\hat{\beta}\,\hat{\rho}$ into the nonminimal coupling (\ref{2.1}), we can see that the whole system is cylindrically symmetric, then, we can work with curvilinear coordinates $x=\rho\,\cos\varphi$ and $y=\rho\,\sin\varphi$. Thereby, we write the line element of the Minkowski spacetime in the form: 
\begin{eqnarray}
ds^{2}=-dt^{2}+d\rho^{2}+\rho^{2}\,d\varphi^{2}+dz^{2}.
\label{2.2a}
\end{eqnarray}

Working with in curvilinear coordinates (both in a background with non-null curvature and with null curvature), the rules of coordinate transformations of spinors obey the rules established in general relativity. This means that the spinor representation of the Lorentz group either can exist or cannot exist under a general coordinate transformation \cite{weinberg,bd}. Therefore, in order to incorporate spinors into a general relativity scenario, one should use the principle of equivalence to define locally inertial frames, where the spinor representation of the Lorentz group is given as in the Minkowski spacetime. In this way, spinors are defined locally by introducing a noncoordinate basis $\hat{\theta}^{a}=e^{a}_{\,\,\,\mu}\left(x\right)\,dx^{\mu}$, whose components $e^{a}_{\,\,\,\mu}\left(x\right)$ are called tetrads and gives rise to the local reference frame of the observers. The tetrads satisfy the following relation: \cite{bd,naka}
\begin{eqnarray}
g_{\mu\nu}\left(x\right)=e^{a}_{\,\,\,\mu}\left(x\right)\,e^{b}_{\,\,\,\nu}\left(x\right)\,\eta_{ab}.
\label{2.3}
\end{eqnarray}
The tetrads also have an inverse defined as $dx^{\mu}=e^{\mu}_{\,\,\,a}\left(x\right)\,\hat{\theta}^{a}$, where 
\begin{eqnarray}
e^{a}_{\,\,\,\mu}\left(x\right)\,e^{\mu}_{\,\,\,b}\left(x\right)=\delta^{a}_{\,\,\,b},\,\,\,\,\,\,\,e^{\mu}_{\,\,\,a}\left(x\right)\,e^{a}_{\,\,\,\nu}\left(x\right)=\delta^{\mu}_{\,\,\,\nu}.
\label{2.4}
\end{eqnarray}

Hence, in curvilinear coordinates (both flat and curved spacetime backgrounds), the relativistic quantum dynamics of a neutral particle with a permanent magnetic dipole moment interacting with external fields is not described by the Dirac equation with the introduction of the nonminimal coupling (\ref{2.1}) anymore. Based on the spinor theory in curved spacetime, the nonminimal coupling (\ref{2.1}) plus the coupling describing the Dirac oscillator (\ref{2}) becomes:
\begin{eqnarray}
i\gamma^{\mu}\,\nabla_{\mu}\rightarrow i\gamma^{\mu}\,\partial_{\mu}+i\gamma^{\mu}\,\Gamma_{\mu}\left(x\right)+i\,\gamma^{\mu}\,m\omega\rho\,\gamma^{0}\,\delta^{\rho}_{\mu}+\frac{\mu}{2}\Sigma^{\mu\nu}\,F_{\mu\nu}\left(x\right),
\label{2.5}
\end{eqnarray}
where $\nabla_{\mu}=\partial_{\mu}+\Gamma_{\mu}\left(x\right)$ corresponds to the components of the covariant derivative of a spinor, with $\Gamma_{\mu}=\frac{i}{4}\,\omega_{\mu ab}\left(x\right)\,\Sigma^{ab}$ being the spinorial connection \cite{bd,naka}, and $\Sigma^{ab}=\frac{i}{2}\left[\gamma^{a},\gamma^{b}\right]$. In the spinor theory in curved spacetime, the $\gamma^{a}$ matrices are defined in the local reference frame of the observers and are identical to the Dirac matrices defined in the Minkowski spacetime (\ref{2.2}). In this notation, the indices $(a,b,c=0,1,2,3)$ indicate the local reference frame, while the indices $\left(\mu,\nu\right)$ indicate the spacetime indices. Thus, the $\gamma^{\mu}$ matrices given in (\ref{2.5}) are related to the $\gamma^{a}$ via $\gamma^{\mu}=e^{\mu}_{\,\,\,a}\left(x\right)\,\gamma^{a}$. Furthermore, the components of the spinorial connection can be obtained by solving the Cartan structure equations \cite{naka} in the absence of torsion: $d\hat{\theta}^{a}+\omega^{a}_{\,\,\,b}\,\hat{\theta}^{b}=0$, where $\omega^{a}_{\,\,\,b}=\omega_{\mu\,\,\,\,b}^{\,\,\,\,a}\left(x\right)\,dx^{\mu}$, and $\omega_{\mu\,\,\,\,b}^{\,\,\,\,a}\left(x\right)$ is called connection 1-form. For instance, we can choose the tetrads for the line element (\ref{2.2a}) being
\begin{eqnarray}
\hat{\theta}^{0}=dt;\,\,\,\,\hat{\theta}^{1}=d\rho;\,\,\,\,\hat{\theta}^{2}=\rho\,d\varphi;\,\,\,\,\hat{\theta}^{3}=dz;
\label{2.6}
\end{eqnarray}

By solving the Cartan structure equations in the absence of torsion, we obtain $\omega_{\varphi\,\,\,2}^{\,\,\,1}\left(x\right)=-\omega_{\varphi\,\,\,1}^{\,\,\,2}\left(x\right)=-1$ and $\gamma^{\mu}\,\Gamma_{\mu}\left(x\right)=\frac{\gamma^{1}}{2\rho}$. Hence, the Dirac equation describing the interaction between the Dirac oscillator and the Aharonov-Casher system is
\begin{eqnarray}
m\Psi=i\gamma^{0}\frac{\partial\Psi}{\partial t}+i\gamma^{1}\left[\frac{\partial}{\partial\rho}+\frac{1}{2\rho}+m\omega\rho\,\gamma^{0}\right]\,\Psi+i\frac{\gamma^{2}}{\eta\rho}\,\frac{\partial\Psi}{\partial\varphi}+i\gamma^{3}\frac{\partial\Psi}{\partial z}+i\mu\vec{\alpha}\cdot\vec{E}\,\Psi-\mu\vec{\Sigma}\cdot\vec{B}\,\Psi.
\label{2.7}
\end{eqnarray}

Moreover, by considering the presence of a radial electric field given by $\vec{E}=\frac{\lambda}{\rho}\,\hat{\rho}$ and the magnetic dipole moment being parallel to the $z$ axis, we can rewrite the Dirac equation (\ref{2.7}) in terms of the Aharonov-Casher geometric phase given in (\ref{1}) in the form: 
\begin{eqnarray}
i\frac{\partial\Psi}{\partial t}=m\hat{\beta}\Psi-i\hat{\alpha}^{1}\left[\frac{\partial}{\partial\rho}+\frac{1}{2\rho}+m\omega\rho\,\hat{\beta}\right]-i\frac{\hat{\alpha}^{2}}{\rho}\frac{\partial\Psi}{\partial\varphi}-i\hat{\alpha}^{3}\,\frac{\partial\Psi}{\partial z}-i\frac{\phi_{\mathrm{AC}}}{2\pi\rho}\,\hat{\beta}\,\hat{\alpha}^{1}\,\Psi.
\label{2.8}
\end{eqnarray}

Hence, the solution of the Dirac equation (\ref{2.8}) is given in the form:
\begin{eqnarray}
\Psi=e^{-i\mathcal{E}t}\,\left(\begin{array}{c}
\phi\\
\xi\\	
\end{array}\right),
\label{2.9}
\end{eqnarray}
where $\phi$ and $\xi$ are spinors of two-components. Then, substituting (\ref{2.9}) into the Dirac equation (\ref{2.8}), we obtain two coupled equation for $\phi$ and $\xi$, where the first coupled equation is
\begin{eqnarray}
\left(\mathcal{E}-m\right)\phi=-i\sigma^{1}\left[\frac{\partial}{\partial\rho}+\frac{1}{2\rho}+\frac{\phi_{\mathrm{AC}}}{2\pi\rho}-m\omega\rho\right]\xi-i\frac{\sigma^{2}}{\rho}\,\frac{\partial\xi}{\partial\varphi}-i\sigma^{3}\frac{\partial\xi}{\partial z},
\label{2.10}
\end{eqnarray}
and the second coupled equation is
\begin{eqnarray}
\left(\mathcal{E}+m\right)\xi=-i\sigma^{1}\left[\frac{\partial}{\partial\rho}+\frac{1}{2\rho}-\frac{\phi_{\mathrm{AC}}}{2\pi\rho}+m\omega\rho\right]\phi-i\frac{\sigma^{2}}{\rho}\,\frac{\partial\phi}{\partial\varphi}-i\sigma^{3}\frac{\partial\phi}{\partial z}.
\label{2.11}
\end{eqnarray}

Eliminating $\xi$ in Eqs. (\ref{2.11}) and (\ref{2.10}), and by considering the dipole moment of the neutral particle is parallel to the $z$ axis of the spacetime, we obtain the following second order differential equation:
\begin{eqnarray}
\left(\mathcal{E}^{2}-m^{2}\right)\phi&=&-\frac{\partial^{2}\phi}{\partial\rho^{2}}-\frac{1}{\rho}\frac{\partial\phi}{\partial\rho}-\frac{1}{\rho^{2}}\,\frac{\partial^{2}\phi}{\partial\varphi^{2}}-\frac{\partial^{2}\phi}{\partial z^{2}}+i\frac{\sigma^{3}}{\rho^{2}}\,\frac{\partial\phi}{\partial\varphi}+\frac{1}{4\rho^{2}}\,\phi\nonumber\\
&+&2i\sigma^{3}\,m\omega\,\frac{\partial\phi}{\partial\varphi}-m\omega\phi-\frac{\phi_{\mathrm{AC}}}{2\pi\rho^{2}}\,\phi-2i\sigma^{3}\frac{\phi_{\mathrm{AC}}}{2\pi\rho^{2}}\frac{\partial\phi}{\partial\varphi}\label{2.12}\\
&+&m^{2}\omega^{2}\rho^{2}\,\phi-2m\omega\,\frac{\phi_{\mathrm{AC}}}{2\pi}\,\phi\,+\left(\frac{\phi_{\mathrm{AC}}}{2\pi}\right)^{2}\frac{\phi}{\rho^{2}}.\nonumber
\end{eqnarray}

We can see in (\ref{2.12}) that $\phi$ is an eigenfunction of the Pauli matrix $\sigma^{3}$, whose eigenvalues are $s=\pm1$, that is, $\sigma^{3}\phi_{s}=s\phi_{s}$, where $\phi_{s}=\left(\phi_{+}\,\,\phi_{-}\right)^{T}$. We also note that the $z$-component of the total angular momentum $\hat{J}_{z}=-i\partial_{\varphi}$ \cite{schu} and the $z$-component of the momentum $\hat{p}_{z}=-i\partial_{z}$ commute with the Hamiltonian of the equation (\ref{2.12}). In that way, we can write the solutions of the equation (\ref{2.12}) in the terms of the eigenvalues of the operators $\hat{p}_{z}=-i\partial_{z}$, and $\hat{J}_{z}=-i\partial_{\varphi}$ \footnote{It has been shown in Ref. \cite{schu} that the $z$-component of the total angular momentum in cylindrical coordinates is given by $\hat{J}_{z}=-i\partial_{\varphi}$, where the eigenvalues are $j=\pm\frac{1}{2},\pm\frac{3}{2},\ldots=l\pm\frac{1}{2}$.}:
\begin{eqnarray}
\phi_{s}\left(\rho,\varphi,z\right)=C\,e^{i\left(l+\frac{1}{2}\right)\varphi}\,e^{ikz}\,R_{s}\left(\rho\right),
\label{2.13}
\end{eqnarray}
where $l=0,\pm1,\pm2,...$, $k$ is a constant and $C$ is a constant. Thus, substituting the solution (\ref{2.12}) into the second order differential equation (\ref{2.12}), we obtain the following radial equation:
\begin{eqnarray}
\frac{d^{2}}{d\rho^{2}}R_{s}\left(\rho\right)+\frac{1}{\rho}\frac{d}{d\rho}R_{s}\left(\rho\right)-\frac{\zeta_{s}^{2}}{\rho^{2}}\,R_{s}\left(\rho\right)-m^{2}\omega^{2}\rho^{2}\,R_{s}\left(\rho\right)+\beta_{s}\,R_{s}\left(\rho\right)=0,
\label{2.14}
\end{eqnarray}
where we have defined in (\ref{2.14}) the parameters:
\begin{eqnarray}
\zeta_{s}&=&l+\frac{1}{2}\left(1-s\right)+s\,\frac{\phi_{\mathrm{AC}}}{2\pi}\nonumber\\
[-2mm]\label{2.15}\\[-2mm]
\beta_{s}&=&\mathcal{E}^{2}-m^{2}-k^{2}+2sm\omega\zeta_{s}+2m\omega.\nonumber
\end{eqnarray}

In order to solve the radial equation (\ref{2.14}), we make a change of variables given by $\tau=m\omega\,\rho^{2}$. Thereby, the radial equation (\ref{2.14}) becomes
\begin{eqnarray}
\left[\tau\,\frac{d^{2}}{d\tau^{2}}+\frac{d}{d\tau}+\left(\frac{\beta_{s}}{4m\omega}-\frac{\zeta_{s}^{2}}{4\tau}-\frac{\tau}{4}\right)\right]R_{s}\left(\tau\right)=0.
\label{2.16}
\end{eqnarray}

We wish to get a solution for Eq. (\ref{2.16}) regular at the origin, then, the solution for Eq. (\ref{2.16}) can be given in the form:
\begin{eqnarray}
R_{s}\left(\tau\right)=e^{-\frac{\tau}{2}}\,\tau^{\frac{\left|\zeta_{s}\right|}{2}}\,F_{s}\left(\tau\right).
\label{2.17}
\end{eqnarray}
In this way, substituting (\ref{2.17}) into (\ref{2.16}), we obtain
\begin{eqnarray}
\tau\frac{d^{2}F_{s}}{d\tau^{2}}+\left(\left|\zeta_{s}\right|+1-\tau\right)\frac{dF_{s}}{d\tau}+\left(\frac{\beta_{s}}{4m\omega}-\frac{\left|\zeta_{s}\right|}{2}-\frac{1}{2}\right)F_{s}=0.
\label{2.18}
\end{eqnarray}

Equation (\ref{2.18}) corresponds to the Kummer equation or the confluent hypergeometric equation \cite{abra}. The function $F_{s}\left(\tau\right)=\,_{1}F_{1}\left(\frac{\left|\zeta_{s}\right|}{2}+\frac{1}{2}-\frac{\beta_{s}}{4m\omega},\,\left|\zeta_{s}\right|+1,\,\tau\right)$ is the Kummer function or the confluent hypergeometric function \cite{abra}. In order to obtain a finite solution everywhere, the confluent hypergeometric series must become a polynomial of degree $n$ \cite{landau}, then, we must impose the parameter $\frac{\left|\zeta_{s}\right|}{2}+\frac{1}{2}-\frac{\beta_{s}}{4m\omega}$ to be equal to a non-positive integer number $-n$ ($n=0,1,2,\ldots$). In this way, by using the relations (\ref{2.15}), we obtain
\begin{eqnarray}
\mathcal{E}^{2}_{n}=m^{2}+k^{2}+4m\omega\left[n+\frac{\left|\zeta_{s}\right|}{2}-s\frac{\zeta_{s}}{2}\right],
\label{2.19}
\end{eqnarray}
where $n=0,1,2,\ldots$. Equation (\ref{2.19}) is the relativistic energy levels of bound states for the Dirac oscillator under the influence of the Aharonov-Casher effect in the Minkowski spacetime. We should note that the relativistic energy levels (\ref{2.19}) depend on the Aharonov-Casher geometric phase $\phi_{\mathrm{AC}}$ with periodicity $\phi_{0}=\pm2\pi$, therefore, we have that $\mathcal{E}_{n,\,l}\left(\phi_{\mathrm{AC}}\pm2\pi\right)=\mathcal{E}_{n,\,l+1}\left(\phi_{\mathrm{AC}}\right)$.

Now, let us obtain the components of the Dirac spinor which are solutions of Eq. (\ref{2.8}). First of all, we must write the expression (\ref{2.13}) in the form:
\begin{eqnarray}
\phi_{s}=C\,e^{i\left(l+\frac{1}{2}-\frac{\sigma^{3}}{2}\right)\varphi}\,e^{ikz}\left(m\omega\right)^{\frac{\left|\zeta_{s}\right|}{2\eta}}\,e^{-m\omega\rho^{2}}\,\rho^{\frac{\left|\zeta_{s}\right|}{\eta}}\,_{1}F_{1}\left(-n,\frac{\left|\zeta_{s}\right|}{\eta}+1;m\omega\rho^{2}\right),
\label{2.21}
\end{eqnarray}
and substitute it into Eq. (\ref{2.11}), with $\xi_{s}=\left(\xi_{+}\,\,\xi_{-}\right)^{T}$. Thus, in order to obtain the positive-energy solutions of the Dirac equation (\ref{2.8}) corresponding to the parallel component to $z$ axis of the spacetime, we must take $s=+1$, and consider $\phi_{-}=0$. In that way, the solution of the Dirac equation (\ref{2.8}) parallel to the $z$ axis of the spacetime is
\begin{eqnarray}
\Psi_{+}&=&f_{+}\,\,_{1}F_{1}\left(-n,\left|\zeta_{+}\right|+1;\frac{\mu\lambda}{2}\rho^{2}\right)\left(
\begin{array}{c}
1	\\
0 \\
\frac{k}{\left(\mathcal{E}+m\right)} \\
\frac{i\,e^{i\varphi}}{\left(\mathcal{E}+m\right)}\left(\mu\lambda\rho-\frac{\left|\zeta_{+}\right|}{\rho}+\frac{\zeta_{+}}{\rho}\right)\\
\end{array}\right)\nonumber\\
[-2mm]\label{2.22}\\[-2mm]
&+&f_{+}\,\frac{i\,e^{i\varphi}}{\left(\mathcal{E}+m\right)}\frac{n\,\mu\lambda\rho}{\left(\left|\zeta_{+}\right|+1\right)}\,\,_{1}F_{1}\left(-n+1,\left|\zeta_{+}\right|+2;\frac{\mu\lambda}{2}\rho^{2}\right)\left(
\begin{array}{c}
0\\
0\\
0\\
1\\	
\end{array}\right).\nonumber
\end{eqnarray}

Next, the positive-energy solutions of the Dirac equation (\ref{2.8}), corresponding to the anti-parallel component to the $z$ axis of the spacetime, is obtained when we choose $s=-1$ and $\phi_{+}=0$. In this way, the solution of the Dirac equation (\ref{2.8}) which is anti-parallel to the $z$ axis is given by
\begin{eqnarray}
\Psi_{-}&=&f_{-}\,\,_{1}F_{1}\left(-n,\left|\zeta_{-}\right|+1;\frac{\mu\lambda}{2}\rho^{2}\right)\left(
\begin{array}{c}
0	\\
1 \\
\frac{i\,e^{-i\varphi}}{\left(\mathcal{E}+m\right)}\left(\mu\lambda\rho-\frac{\left|\zeta_{-}\right|}{\rho}-\frac{\zeta_{-}}{\rho}\right)\\
-\frac{k}{\left(\mathcal{E}+m\right)} \\
\end{array}\right)\nonumber\\
[-2mm]\label{2.23}\\[-2mm]
&+&f_{-}\,\frac{i\,e^{-i\varphi}}{\left(\mathcal{E}+m\right)}\frac{\,n\,\mu\lambda\rho}{\left(\left|\zeta_{-}\right|+1\right)}\,\,_{1}F_{1}\left(-n+1,\left|\zeta_{-}\right|+2;\frac{\mu\lambda}{2}\rho^{2}\right)\left(
\begin{array}{c}
0\\
0\\
1\\
0\\	
\end{array}\right),\nonumber
\end{eqnarray}
where we have defined the parameters $f_{s}$ in (\ref{2.22}) and (\ref{2.23}) as
\begin{eqnarray}
f_{\pm}=f_{s}=C\,e^{-i\mathcal{E}t}\,e^{i\left(l+\frac{1}{2}-\frac{s}{2}\right)\varphi}\,e^{ikz}\left(\frac{\mu\lambda}{2}\right)^{\frac{\left|\zeta_{s}\right|}{2}}\,e^{-\frac{\mu\lambda}{4}\rho^{2}}\,\rho^{\left|\zeta_{s}\right|}.
\label{2.24}
\end{eqnarray} 

Hence, the components of the Dirac spinor (\ref{2.22}) and (\ref{2.23}) correspond to the positive-energy solutions of the Dirac equation (\ref{2.8}). Moreover, we can also obtain the negative-energy solutions of the Dirac equation (\ref{2.8}) by applying the same procedure used to obtain Eqs. (\ref{2.22}) and (\ref{2.23}).

\section{the Dirac oscillator and the relativistic Aharonov-Casher system in the cosmic string spacetime}
 
In this section, we discuss the influence of the Aharonov-Casher effect \cite{ac} and curvature effects on the Dirac oscillator. This study is made by considering a topological defect background given by the cosmic string spacetime. We begin this section by introducing the cosmic string spacetime background and, by using the tetrad formalism discussed in the previous section, we solve the Dirac equation that describes the interaction between the Dirac oscillator and the Aharonov-Casher system \cite{ac} in the cosmic string spacetime. 

The study of the influence of topological defects on physical systems has been emphasized in several works \cite{vil,kibble,kleinert,anp,kat,furt,moraesG,moraesG2,moraesG3,b6}. Recently, relativistic \cite{jac,vg} and nonrelativistic \cite{furt3,baus,azev} quantum dynamics of spinless particles has been studied in in the presence of topological defects. The Landau quantization has also been studied in the presence of a topological defect in the nonrelativistic case by showing that the degeneracy of the Landau levels is broken due to the presence of the topological defect \cite{furt4,furt5,furt6}. Thus, the cosmic string spacetime is described by the following line element:
\begin{eqnarray}
ds^{2}=-dt^{2}+d\rho^{2}+\eta^{2}\rho^{2}d\varphi^{2}+dz^{2}.
\label{3.1}
\end{eqnarray}
where the parameter $\eta$ is related to the deficit angle, which is defined as $\eta=1-4\varpi G/c^{2}$ ($\varpi$ being the linear mass density of the cosmic string). The azimuthal angle is defined in the range $0\leq\varphi<2\pi$, while the parameter related to the deficit angle is defined in the range $0\,<\,\eta\,<\,1$. For all values where $\eta\,>\,1$, the cosmic string spacetime is not defined anymore because this case corresponds to a spacetime with negative curvature. Topological defects with negative curvature does make sense only in the description of linear topological defects in crystalline solids \cite{kat,furt}. The geometry described by the line element (\ref{3.1}) possesses a conical singularity \cite{staro} that gives rise to the curvature concentrated on the cosmic string axis. This conical singularity is represented by the curvature tensor:
\begin{eqnarray}
R_{\rho,\varphi}^{\rho,\varphi}=\frac{1-\eta}{4\eta}\,\delta_{2}(\vec{r}),
\label{3.2}
\end{eqnarray}
where $\delta_{2}(\vec{r})$ is the two-dimensional delta function. As we have discussed in the previous section, in a background with non-null curvature, we can work with the spinor theory in curved spacetime \cite{weinberg,bd} in order to incorporate spinors into the general relativity scenario. By following the steps from Eq. (\ref{2.3}) to Eq. (\ref{2.5}), we can define the tetrads and the inverse of the tetrads for the metric describing the cosmic string spacetime (\ref{3.1}) as it follows \cite{bf11}:
\begin{eqnarray}
e^{a}_{\,\,\,\mu}\left(x\right)=\left(
\begin{array}{cccc}
1 & 0 & 0 & 0 \\
0 & \cos\varphi & -\eta\rho\sin\varphi & 0 \\
0 & \sin\varphi & \eta\rho\cos\varphi & 0 \\
0 & 0 & 0 & 1 \\
\end{array}\right),\,\,\,\,\,\,\,\,\,\,\,\,\,e^{\mu}_{\,\,\,a}\left(x\right)=\left(
\begin{array}{cccc}
1 & 0 & 0 & 0 \\
0 & \cos\varphi & \sin\varphi & 0 \\
0 & -\frac{\sin\varphi}{\eta\rho} & \frac{\cos\varphi}{\eta\rho} & 0 \\
0 & 0 & 0 & 1 \\
\end{array}\right).
\label{3.3}
\end{eqnarray}

By taking the tetrads (\ref{3.3}) and solving the Cartan structure equations in the absence of torsion, we obtain the following non-null components of the connection 1-form: $\omega_{\varphi\,\,\,2}^{\,\,\,1}\left(x\right)=-\omega_{\varphi\,\,\,1}^{\,\,\,2}\left(x\right)=1-\eta$. Substituting $\omega_{\varphi\,\,\,2}^{\,\,\,1}\left(x\right)=-\omega_{\varphi\,\,\,1}^{\,\,\,2}\left(x\right)=\left(1-\eta\right)$ into the definition of the spinorial connection $\Gamma_{\mu}\left(x\right)$, we obtain
\begin{eqnarray}
\Gamma_{\varphi}\left(x\right)=\frac{i}{2}\left(1-\eta\right)\,\Sigma^{3}.
\label{3.4}
\end{eqnarray}

Hence, the Dirac equation describing the interaction between the Dirac oscillator and the Aharonov-Casher system in the cosmic string spacetime becomes
\begin{eqnarray}
m\Psi&=&i\gamma^{t}\frac{\partial\Psi}{\partial t}+i\gamma^{\rho}\left[\frac{\partial}{\partial\rho}-\frac{1}{2}\frac{\left(1-\eta\right)}{\eta\rho}+m\omega\rho\,\gamma^{0}\right]\,\Psi+i\frac{\gamma^{\varphi}}{\eta\rho}\,\frac{\partial\Psi}{\partial\varphi}+i\gamma^{z}\frac{\partial\Psi}{\partial z}\nonumber\\
[-2mm]\label{3.5}\\[-2mm]
&+&i\mu\vec{\alpha}\cdot\vec{E}\,\Psi-\mu\vec{\Sigma}\cdot\vec{B}\,\Psi.\nonumber
\end{eqnarray}

By considering a linear distribution of electric charges on the $z$ axis, it has been shown in Ref. \cite{moraesG2} that the topology of the cosmic string changes the electric field resulting in $\vec{E}=\frac{\lambda}{\eta\rho}\,\hat{\rho}$. Based on this result, the influence of the topology of the cosmic string on the Aharonov-Casher effect \cite{ac} has been studied in \cite{bf1}. In that way, the Dirac equation describing the interaction between the Dirac oscillator and the relativistic Aharonov-Casher system in the cosmic string spacetime is   
\begin{eqnarray}
m\Psi=i\gamma^{t}\frac{\partial\Psi}{\partial t}+i\gamma^{\rho}\left[\frac{\partial}{\partial\rho}-\frac{1}{2}\frac{\left(1-\eta\right)}{\eta\rho}-\frac{\phi_{\mathrm{AC}}}{2\pi\eta\rho}\,\gamma^{0}+m\omega\rho\,\gamma^{0}\right]\,\Psi+i\frac{\gamma^{\varphi}}{\eta\rho}\,\frac{\partial\Psi}{\partial\varphi}+i\gamma^{z}\frac{\partial\Psi}{\partial z},
\label{3.6}
\end{eqnarray}
where we have used the tetrads (\ref{3.3}), and defined the $\gamma^{\mu}$ matrices in Eqs. (\ref{3.5}) and (\ref{3.6}) in the form \cite{bf1}:
\begin{eqnarray}
\gamma^{t}&=&\gamma^{0}\nonumber\\
\gamma^{\rho}&=&\cos\varphi\,\gamma^{1}+\sin\varphi\,\gamma^{2}\nonumber\\
[-4mm]\label{3.7}\\[-4mm]
\gamma^{\varphi}&=&-\sin\varphi\,\gamma^{1}+\cos\varphi\,\gamma^{2}\nonumber\\
\gamma^{z}&=&\gamma^{3}.\nonumber
\end{eqnarray}

The solution of the Dirac equation (\ref{3.6}) is given in the same form as given in (\ref{2.9}). Substituting (\ref{2.9}) into (\ref{3.6}), we also obtain two coupled equations for $\phi$ and $\xi$, where the first coupled equation is
\begin{eqnarray}
\left(\mathcal{E}-m\right)\phi=-i\sigma^{\rho}\left[\frac{\partial}{\partial\rho}-\frac{1}{2}\frac{\left(1-\eta\right)}{\eta\rho}+\frac{\phi_{\mathrm{AC}}}{2\pi\eta\rho}-m\omega\rho\right]\xi-i\frac{\sigma^{\varphi}}{\eta\rho}\,\frac{\partial\xi}{\partial\varphi}-i\sigma^{z}\frac{\partial\xi}{\partial z},
\label{3.8}
\end{eqnarray}
while the second coupled equation is
\begin{eqnarray}
\left(\mathcal{E}+m\right)\xi=-i\sigma^{\rho}\left[\frac{\partial}{\partial\rho}-\frac{1}{2}\frac{\left(1-\eta\right)}{\eta\rho}-\frac{\phi_{\mathrm{AC}}}{2\pi\eta\rho}+m\omega\rho\right]\phi-i\frac{\sigma^{\varphi}}{\eta\rho}\,\frac{\partial\phi}{\partial\varphi}-i\sigma^{z}\frac{\partial\phi}{\partial z}
\label{3.9}
\end{eqnarray}
where we have defined the $\vec{\sigma}$ matrices in (\ref{3.8}) and ({\ref{3.9}) as 
\begin{eqnarray}
\sigma^{\rho}=\cos\varphi\sigma^{1}+\sin\varphi\sigma^{2}\,;\,\,\,\sigma^{\varphi}=-\sin\varphi\sigma^{1}+\cos\varphi\sigma^{2}\,;\,\,\,\sigma^{z}=\sigma^{3}. 
\label{3.10}
\end{eqnarray}

Eliminating $\xi$ in Eqs. (\ref{3.10}) and (\ref{3.9}) and by considering the dipole moment of the neutral particle being parallel to the $z$ axis of the spacetime, we have
\begin{eqnarray}
\left(\mathcal{E}^{2}-m^{2}\right)\phi&=&-\frac{\partial^{2}\phi}{\partial\rho^{2}}-\frac{1}{\rho}\frac{\partial\phi}{\partial\rho}-\frac{1}{\eta^{2}\rho^{2}}\,\frac{\partial^{2}\phi}{\partial\varphi}-\frac{\partial^{2}\phi}{\partial z^{2}}-i\sigma^{3}\frac{\left(1-\eta\right)}{\eta^{2}\rho^{2}}\,\frac{\partial\phi}{\partial\varphi}\nonumber\\
&-&2i\sigma^{3}\,\frac{\phi_{\mathrm{AC}}}{2\pi\eta^{2}\rho^{2}}\,\frac{\partial\phi}{\partial\varphi}+\frac{1}{4}\frac{\left(1-\eta\right)^{2}}{\eta^{2}\rho^{2}}\phi-2i\sigma^{3}\frac{m\omega}{\eta}\,\frac{\partial\phi}{\partial\varphi}+\frac{m\omega}{\eta}\phi \label{3.11}\\ 
&+&\frac{\phi_{\mathrm{AC}}}{2\pi}\frac{\left(1-\eta\right)}{\eta^{2}\rho^{2}}\phi+\left(\frac{\phi_{\mathrm{AC}}}{2\pi\eta\rho}\right)^{2}\phi+2\frac{m\omega\,\phi_{\mathrm{AC}}}{2\pi\eta}\phi+m\omega\phi+m^{2}\omega^{2}\rho^{2}\phi.\nonumber
\end{eqnarray}

We can also see in Eq. (\ref{3.11}) that $\phi$ is an eigenfunction of the Pauli matrix $\sigma^{3}$, whose eigenvalues are $s=\pm1$, that is, $\sigma^{3}\phi_{s}=s\phi_{s}$, where $\phi_{s}=\left(\phi_{+}\,\,\phi_{-}\right)^{T}$. We also note that the $z$-component of the total angular momentum can also be written in the form: $\hat{J}_{z}=-i\partial_{\varphi}+\frac{\sigma^{3}}{2}$ \cite{mello}. Besides, we can also see that both operators $\hat{J}_{z}$ and $\hat{p}_{z}=-i\partial_{z}$ commute with the Hamiltonian of the equation (\ref{3.11}). In that way, we can write the solutions of the equation (\ref{3.11}) in the terms of the eigenvalues of the operators $\hat{p}_{z}=-i\partial_{z}$, and $\hat{J}_{z}=-i\partial_{\varphi}$, that is, 
\begin{eqnarray}
\phi_{s}\left(\rho,\varphi,z\right)=C\,e^{i\left(l+\frac{1}{2}-\frac{\sigma^{3}}{2}\right)\varphi}\,e^{ikz}\,R_{s}\left(\rho\right)
\label{3.12}
\end{eqnarray}
where $l=0,\pm1,\pm2,...$, $k$ is a constant and $C$ is a constant. Substituting the solution (\ref{3.12}) into (\ref{3.11}), we obtain the following radial equation:
\begin{eqnarray}
\frac{d^{2}}{d\rho^{2}}R_{s}\left(\rho\right)&+&\frac{1}{\rho}\frac{d}{d\rho}R_{s}\left(\rho\right)-\frac{\tilde{\zeta}_{s}^{2}}{\eta^{2}\rho^{2}}R_{s}\left(\rho\right)-m^{2}\omega^{2}\rho^{2}\,R_{s}\left(\rho\right)+\beta_{s}\,R_{s}\left(\rho\right)=0,
\label{3.13}
\end{eqnarray}
where we have defined in (\ref{3.12}) the parameters:  
\begin{eqnarray}
\tilde{\zeta}_{s}&=&l+\frac{1}{2}\left(1-s\right)+s\,\frac{\phi_{\mathrm{AC}}}{2\pi}+\frac{s}{2}\left(1-\eta\right)\nonumber\\
[-2mm]\label{3.14}\\[-2mm]
\tilde{\beta}_{s}&=&\mathcal{E}^{2}-m^{2}-k^{2}-2s\frac{m\omega}{\eta}\,\tilde{\zeta}_{s}+2m\omega.\nonumber
\end{eqnarray}

Therefore, in order to solve the radial equation (\ref{3.13}), we make a change of variables $\tau=m\omega\,\rho^{2}$ and rewrite Eq. (\ref{3.13}) in the form:
\begin{eqnarray}
\left[\tau\,\frac{d^{2}}{d\tau^{2}}+\frac{d}{d\tau}+\left(\frac{\tilde{\beta}_{s}}{4m\omega}-\frac{\tilde{\zeta}_{s}^{2}}{4\eta^{2}\tau}-\frac{\tau}{4}\right)\right]R_{s}\left(\tau\right)=0.
\label{3.15}
\end{eqnarray}

As discussed in the previous section, we need a solution regular at the origin, then, the solution for Eq. (\ref{3.15}) has the following form:
\begin{eqnarray}
R_{s}\left(\tau\right)=e^{-\frac{\tau}{2}}\,\tau^{\frac{\left|\tilde{\zeta}_{s}\right|}{2\eta}}\,F_{s}\left(\tau\right).
\label{3.16}
\end{eqnarray}
In this way, substituting (\ref{3.16}) into (\ref{3.15}), we obtain the following second order differential equation:
\begin{eqnarray}
\tau\frac{d^{2}F_{s}}{d\tau^{2}}+\left(\frac{\left|\tilde{\zeta}_{s}\right|}{\eta}+1-\tau\right)\frac{dF_{s}}{d\tau}+\left(\frac{\tilde{\beta}_{s}}{4m\omega}-\frac{\left|\tilde{\zeta}_{s}\right|}{2\eta}-\frac{1}{2}\right)F_{s}=0,
\label{3.17}
\end{eqnarray}
which corresponds to the Kummer equation or confluent hypergeometric equation \cite{abra}. We also have that the function $F_{s}\left(\tau\right)=\,_{1}F_{1}\left(\frac{\left|\tilde{\zeta}_{s}\right|}{2\eta}+\frac{1}{2}-\frac{\tilde{\beta}_{s}}{4m\omega},\frac{\left|\tilde{\zeta}_{s}\right|}{\eta}+1;\tau\right)$ corresponds to the Kummer function of first kind \cite{abra}. Therefore, in order to obtain a finite solution at each region of the spacetime, we impose the condition where the parameter $\frac{\left|\tilde{\zeta}_{s}\right|}{2\eta}+\frac{1}{2}-\frac{\tilde{\beta}_{s}}{4m\omega}$ is equal to a non-positive number. Hence, by taking $\frac{\left|\tilde{\zeta}_{s}\right|}{2\eta}+\frac{1}{2}-\frac{\tilde{\beta}_{s}}{4m\omega}=-n$, where $n=0,1,2,\ldots$, we obtain
\begin{eqnarray}
\mathcal{E}^{2}_{n}=m^{2}+k^{2}+4m\omega\left[n+\frac{\left|\tilde{\zeta}_{s}\right|}{2\eta}-s\frac{\tilde{\zeta}_{s}}{2\eta}\right],
\label{3.18}
\end{eqnarray}
which are the relativistic energy levels for bound states for a Dirac oscillator under the influence of the Aharonov-Casher effect in the cosmic string spacetime. Again, we can note that the relativistic energy levels depend on the Aharonov-Casher geometric phase $\phi_{\mathrm{AC}}$. However, comparing with the results obtained in the previous section (\ref{2.19}), we can see that the curvature of the conical surface breaks the degeneracy of the relativistic energy levels of the Dirac oscillator under the influence of the Aharonov-Casher effect in the same way of the breaking of the degeneracy of the relativistic Landau quantization for a scalar quantum particle \cite{furt7} and for neutral particles \cite{bf11}. 

Observe that the presence of the Aharonov-Casher geometric phase $\phi_{\mathrm{AC}}$ in the relativistic energy levels (\ref{3.18}) shows that the relativistic spectrum of energy has a periodicity $\phi_{0}=\pm2\pi$, then, we have that $\mathcal{E}_{n,\,l}\left(\phi_{\mathrm{AC}}\pm2\pi\right)=\mathcal{E}_{n,\,l+1}\left(\phi_{\mathrm{AC}}\right)$. In contrast to the result obtained in Eq. (\ref{2.19}), the effects of curvature on the relativistic energy levels (\ref{3.18}) change the pattern of oscillations of the spectrum of energy. However, by taking the limit $\eta\rightarrow1$, we can observe that we recover all results given in the Minkowski spacetime obtained in the previous section.

Now, by following the steps from Eq. (\ref{2.21}) to Eq. (\ref{2.24}), we can obtain the components of the Dirac spinor that are parallel and anti-parallel to the $z$ axis. First of all, we must write the expression (\ref{3.12}) in the form
\begin{eqnarray}
\phi_{s}=C\,e^{i\left(l+\frac{1}{2}-\frac{\sigma^{3}}{2}\right)\varphi}\,e^{ikz}\left(m\omega\right)^{\frac{\left|\tilde{\zeta}_{s}\right|}{2\eta}}\,e^{-m\omega\rho^{2}}\,\rho^{\frac{\left|\tilde{\zeta}_{s}\right|}{\eta}}\,_{1}F_{1}\left(-n,\frac{\left|\tilde{\zeta}_{s}\right|}{\eta}+1;m\omega\rho^{2}\right).
\label{3.20}
\end{eqnarray}

In this way, the component of the positive-energy solution of the Dirac equation (\ref{3.6}) which is parallel to the $z$-axis is
\begin{eqnarray}
\Psi_{+}&=&g_{+}\,\,_{1}F_{1}\left(-n,\frac{\left|\tilde{\zeta}_{+}\right|}{\eta}+1;\frac{\mu\lambda}{2}\rho^{2}\right)\left(
\begin{array}{c}
1	\\
0 \\
\frac{k}{\left(\mathcal{E}+m\right)} \\
\frac{i\,e^{i\varphi}}{\left(\mathcal{E}+m\right)}\left(\mu\lambda\rho-\frac{\left|\tilde{\zeta}_{+}\right|}{\eta\rho}+\frac{\tilde{\zeta}_{+}}{\eta\rho}\right)\\
\end{array}\right)\nonumber\\
[-2mm]\label{3.21}\\[-2mm]
&+&g_{+}\,\frac{i\,e^{i\varphi}}{\left(\mathcal{E}+m\right)}\frac{n\,\mu\lambda\rho}{\left(\frac{\left|\tilde{\zeta}_{+}\right|}{\eta}+1\right)}\,_{1}F_{1}\left(-n+1,\frac{\left|\tilde{\zeta}_{+}\right|}{\eta}+2;\frac{\mu\lambda}{2}\rho^{2}\right)\left(
\begin{array}{c}
0\\
0\\
0\\
1\\	
\end{array}\right),\nonumber
\end{eqnarray}
and, the component of positive-energy solution of the Dirac equation (\ref{3.6}) which is anti-parallel to the $z$ axis is
\begin{eqnarray}
\Psi_{-}&=&g_{-}\,\,_{1}F_{1}\left(-n,\frac{\left|\tilde{\zeta}_{-}\right|}{\eta}+1;\frac{\mu\lambda}{2}\rho^{2}\right)\left(
\begin{array}{c}
0	\\
1 \\
\frac{i\,e^{-i\varphi}}{\left(\mathcal{E}+m\right)}\left(\mu\lambda\rho-\frac{\left|\tilde{\zeta}_{-}\right|}{\eta\rho}-\frac{\tilde{\zeta}_{-}}{\eta\rho}\right)\\
-\frac{k}{\left(\mathcal{E}+m\right)} \\
\end{array}\right)\nonumber\\
[-2mm]\label{3.22}\\[-2mm]
&+&g_{-}\,\frac{i\,e^{-i\varphi}}{\left(\mathcal{E}+m\right)}\frac{\,n\,\mu\lambda\rho}{\left(\frac{\left|\tilde{\zeta}_{-}\right|}{\eta}+1\right)}\,_{1}F_{1}\left(-n+1,\frac{\left|\tilde{\zeta}_{-}\right|}{\eta}+2;\frac{\mu\lambda}{2}\rho^{2}\right)\left(
\begin{array}{c}
0\\
0\\
1\\
0\\	
\end{array}\right),\nonumber
\end{eqnarray}
where we have defined the parameter $g_{\pm}=g_{s}$ in (\ref{3.21}) and (\ref{3.22}) as
\begin{eqnarray}
g_{s}=C\,e^{-i\mathcal{E}t}\,e^{i\left(l+\frac{1}{2}-\frac{s}{2}\right)\varphi}\,e^{ikz}\left(\frac{\mu\lambda}{2}\right)^{\frac{\left|\tilde{\zeta}_{s}\right|}{2\eta}}\,e^{-\frac{\mu\lambda}{4}\rho^{2}}\,\rho^{\left|\tilde{\zeta}_{s}\right|}.
\label{3.23}
\end{eqnarray} 

Hence, we have obtained in Eqs. (\ref{3.21}) and (\ref{3.22}) the components of positive-energy solutions of the Dirac equation (\ref{3.6}). Negative-energy solutions of the Dirac equation (\ref{3.6}) can also be obtained by applying the same procedure used to obtain Eqs. (\ref{3.21}) and (\ref{3.22}) as we have discussed in this section, and in the previous section.

\section{the Dirac oscillator and the relativistic Aharonov-Casher system in the cosmic dislocation spacetime}

In this section, we consider a spacetime background with the presence of torsion and curvature. Hence, we study the influence of the relativistic Aharonov-Casher system \cite{ac} on the Dirac oscillator in the cosmic dislocation background. The cosmic dislocation background is described by the following line element: 
\begin{eqnarray}
ds^{2}=-dt^{2}+d\rho^{2}+\eta^{2}\rho^{2}d\varphi^{2}+\left(dz+\chi d\varphi\right)^{2}.
\label{4.1}
\end{eqnarray}

The parameter $\chi$ is related to the torsion of the defect or, by using the crystallography language, the parameter $\chi$ is related to the Burgers vector. The parameter $\eta$ is related to the deficit angle as we have shown in the previous section. In the same way of the cosmic string spacetime, the cosmic dislocation spacetime has a non-null curvature tensor that gives rise to the conical singularity \cite{staro}. The curvature tensor of the cosmic dislocation spacetime is equal to the cosmic string curvature tensor given in (\ref{3.2}).  

In the presence of torsion and curvature, it is also convenient to work the Dirac equation by using the spinor theory in curved spacetime \cite{bd,weinberg}. By following the steps from Eq. (\ref{2.3}) to Eq. (\ref{2.6}), we can define the local reference frame of the observers for the metric (\ref{4.1}) in the form \cite{bf2}:
\begin{eqnarray}
e^{a}_{\,\,\,\mu}\left(x\right)=\left(
\begin{array}{cccc}
1 & 0 & 0 & 0 \\
0 & \cos\varphi & -\eta\rho\sin\varphi & 0 \\
0 & \sin\varphi & \eta\rho\cos\varphi & 0 \\
0 & 0 & \chi & 1 \\
\end{array}\right),\,\,\,\,\,\,\,\,\,\,\,\,\,e^{\mu}_{\,\,\,a}\left(x\right)=\left(
\begin{array}{cccc}
1 & 0 & 0 & 0 \\
0 & \cos\varphi & \sin\varphi & 0 \\
0 & -\frac{\sin\varphi}{\eta\rho} & \frac{\cos\varphi}{\eta\rho} & 0 \\
0 & \frac{\chi}{\eta\rho}\sin\varphi & -\frac{\chi}{\eta\rho}\cos\varphi & 1 \\
\end{array}\right).
\label{4.2}
\end{eqnarray}

From now on, we need to solve the Cartan structure equations in the presence of torsion \cite{naka}: $T^{a}=d\hat{\theta}^{a}+\omega^{a}_{\,\,\,b}\,\hat{\theta}^{b}$, where $T^{a}=T^{a}_{\,\,\,\mu\nu}\left(x\right)\,dx^{\mu}\,dx^{\nu}$ corresponds to torsion 2-forms and $\omega^{a}_{\,\,\,b}=\omega_{\mu\,\,\,\,b}^{\,\,\,\,a}\left(x\right)\,dx^{\mu}$, with $\omega_{\mu\,\,\,\,b}^{\,\,\,\,a}\left(x\right)$ being the connection 1-form as we have discussed in the previous section. By solving the Cartan structure equation for the tetrads given in (\ref{4.2}), we have
\begin{eqnarray}
T^{3}=2\pi\chi\,\delta\left(\rho\right)\,\delta\left(\varphi\right)\,d\varphi\wedge d\rho\,;\,\,\,\,\,\,\,\,\,\,\omega_{\varphi\,\,\,2}^{\,\,\,1}\left(x\right)=-\omega_{\varphi\,\,\,1}^{\,\,\,2}\left(x\right)=1-\eta,
\label{4.3}
\end{eqnarray}
where we can see the presence of a non-null component of the torsion 2-forms. From the spinor theory in curved space \cite{weinberg}, the presence of torsion in the spacetime results in the following expression for the components of the covariant derivative of a spinor \cite{shap}: 
\begin{eqnarray}
\bar{\nabla}_{\mu}=\partial_{\mu}+\Gamma_{\mu}\left(x\right)+K_{\mu}\left(x\right),
\label{4.4}
\end{eqnarray}
where $\Gamma_{\mu}\left(x\right)=\frac{i}{4}\,\omega_{\mu ab}\left(x\right)\,\Sigma^{ab}$ is the spinorial connection discussed previously \cite{bd,naka} and $K_{\mu}\left(x\right)=\frac{i}{4}K_{\mu ab}\left(x\right)\Sigma^{ab}$. The connection $1$-form $K_{\mu ab}\left(x\right)$ is related to the contortion tensor by \cite{shap}:
\begin{eqnarray}
K_{\mu ab}\left(x\right)= K_{\beta\nu\mu}\left[e^{\nu}_{\,\,\,a}\left(x\right)\,e^{\beta}_{\,\,\,b}\left(x\right)-e^{\nu}_{\,\,\,b}\left(x\right)\,e^{\beta}_{\,\,\,a}\left(x\right)\right].
\label{4.5}
\end{eqnarray}

By following the definitions of Ref. \cite{shap}, the contortion tensor is related to the torsion tensor via $K^{\beta}_{\,\,\,\nu\mu}=\frac{1}{2}\left(T^{\beta}_{\,\,\,\nu\mu}-T_{\nu\,\,\,\,\mu}^{\,\,\,\beta}-T^{\,\,\,\beta}_{\mu\,\,\,\,\nu}\right)$, where we have that the torsion tensor is antisymmetric in the last two indices, while the contortion tensor is antisymmetric in the first two indices. Moreover, it is usually convenient to write the torsion tensor into three irreducible components: the trace vector $\bar{T}_{\mu}=T^{\beta}_{\,\,\,\mu\beta}$, the axial vector $S^{\alpha}=\epsilon^{\alpha\beta\nu\mu}\,T_{\beta\nu\mu}$ and the tensor $q_{\beta\nu\mu}$, which satisfies the conditions $q^{\beta}_{\,\,\mu\beta}=0$ and $\epsilon^{\alpha\beta\nu\mu}\,q_{\beta\nu\mu}=0$. Thus, the torsion tensor becomes
\begin{eqnarray}
T_{\beta\nu\mu}=\frac{1}{3}\left(\bar{T}_{\nu}\,g_{\beta\mu}-\bar{T}_{\mu}\,g_{\beta\nu}\right)-\frac{1}{6}\,\epsilon_{\beta\nu\mu\gamma}\,S^{\gamma}+q_{\beta\nu\mu}.
\label{4.6}
\end{eqnarray}

As it has been shown in Ref. \cite{shap}, the trace vector $\bar{T}_{\mu}$ and the tensor $q_{\beta\nu\mu}$ decouples with fermions, then, by introducing the nonminimal coupling (\ref{2.5}) into the Dirac equation (where the components of the covariant derivative is given by (\ref{4.4})), the Dirac equation that describes the interaction between the Dirac oscillator and the relativistic Aharonov-Casher system in the cosmic dislocation spacetime is given by
\begin{eqnarray}
m\Psi&=&i\gamma^{t}\frac{\partial\Psi}{\partial t}+i\gamma^{\rho}\left[\frac{\partial}{\partial\rho}-\frac{1}{2}\frac{\left(1-\eta\right)}{\eta\rho}+m\omega\rho\,\gamma^{0}\right]\,\Psi+i\frac{\gamma^{\varphi}}{\eta\rho}\left[\frac{\partial}{\partial\varphi}-\chi\frac{\partial}{\partial z}\right]\,\Psi+i\gamma^{z}\frac{\partial\Psi}{\partial z}\nonumber\\
[-2mm]\label{4.7}\\[-2mm]
&+&\frac{1}{8}\,S^{0}\gamma^{t}\gamma^{5}\,\Psi-\frac{1}{8}\,\vec{\Sigma}\cdot\vec{S}\,\Psi+\mu\,\vec{\alpha}\cdot\vec{E}\,\Psi-\mu\,\vec{\Sigma}\cdot\vec{B}\,\Psi,\nonumber
\end{eqnarray}
where we have used the tetrads (\ref{4.2}), and the results of (\ref{4.3}) to obtain $\Gamma_{\varphi}\left(x\right)=\frac{i}{2}\left(1-\eta\right)\,\Sigma^{3}$. We have also defined the $\gamma^{\mu}$ matrices in (\ref{4.7}) in the same way of (\ref{3.7}). From the definition of the the axial vector $S^{\alpha}$ given above, and from the results (\ref{4.3}), one can check that the only non-null component of the axial vector is $S^{0}=-\frac{4\pi\chi}{\eta\rho}\,\delta\left(\rho\right)\,\delta\left(\varphi\right)$ \cite{bf2}. Moreover, it has been shown in Ref. \cite{moraesG3} that the electromagnetic field can be modified by the presence of torsion. However, for a radial electric field $\vec{E}=\frac{\lambda}{\eta\rho}\,\hat{\rho}$, there is no influence of the torsion described by the line element (\ref{4.1}) \cite{moraesG3,bf2}. Thus, we can write the Dirac equation (\ref{4.7}) in the form:
\begin{eqnarray}
m\Psi=i\gamma^{t}\frac{\partial\Psi}{\partial t}+i\gamma^{\rho}\left[\frac{\partial}{\partial\rho}-\frac{1}{2}\frac{\left(1-\eta\right)}{\eta\rho}-\frac{\mu\lambda}{\eta\rho}\,\gamma^{0}+m\omega\rho\,\gamma^{0}\right]\,\Psi+i\frac{\gamma^{\varphi}}{\eta\rho}\left[\frac{\partial}{\partial\varphi}-\chi\frac{\partial}{\partial z}\right]\,\Psi+i\gamma^{z}\frac{\partial\Psi}{\partial z}.
\label{4.8}
\end{eqnarray}

By taking the solution for the Dirac equation (\ref{4.8}) in the same form of Eq. (\ref{2.9}), and by following the steps from Eq. (\ref{3.8}) to Eq. (\ref{3.18}), we obtain
\begin{eqnarray}
\mathcal{E}^{2}_{n}=m^{2}+k^{2}+4m\omega\left[n+\frac{\left|\bar{\zeta}_{s}\right|}{2\eta}-s\frac{\bar{\zeta}_{s}}{2\eta}\right],
\label{4.9}
\end{eqnarray}
where $n=0,1,2,\ldots$, $k$ is a constant, $l=0,\pm1,\pm2,\ldots$ and 
\begin{eqnarray}
\bar{\zeta}_{s}=l+\frac{1}{2}\left(1-s\right)+\frac{s}{2}\left(1-\eta\right)-\chi k+s\frac{\phi_{\mathrm{AC}}}{2\pi}.
\label{eq:}
\end{eqnarray}

Equation (\ref{4.9}) corresponds to the relativistic energy levels for the Dirac oscillator under the influence of the Aharonov-Casher effect in the cosmic dislocation spacetime. We can see in Eq. (\ref{4.9}) that both torsion and curvature breaks the degeneracy of the relativistic energy levels obtained in (\ref{2.19}) as discussed in the previous section. Observe that, by taking the limit $\eta\rightarrow1$, we have only the influence of torsion on the relativistic energy levels (\ref{4.9}), which is given by
\begin{eqnarray}
\mathcal{E}^{2}_{n}=m^{2}+k^{2}+4m\omega\left[n+\frac{\left|l+\frac{1}{2}\left(1-s\right)-\chi k+s\frac{\phi_{\mathrm{AC}}}{2\pi}\right|}{2}-s\,\frac{\left(l+\frac{1}{2}\left(1-s\right)-\chi k+s\frac{\phi_{\mathrm{AC}}}{2\pi}\right)}{2}\right].
\label{4.9a}
\end{eqnarray}
We can also note that, by taking the limit $\eta\rightarrow1$ and $\chi=0$, we recover the relativistic energy levels in the Minkowski spacetime given in (\ref{2.19}). 

We can also see that the relativistic spectrum of energy (\ref{4.9}) is a periodic function of the Aharonov-Casher geometric phase, with periodicity $\phi_{0}=\pm2\pi$. Thereby, we also have that $\mathcal{E}_{n,\,l}\left(\phi_{\mathrm{AC}}\pm2\pi\right)=\mathcal{E}_{n,\,l+1}\left(\phi_{\mathrm{AC}}\right)$. In this case, we have that the effect of both torsion and curvature on the relativistic energy levels (\ref{4.9a}) change the pattern of oscillations of the spectrum of energy. Besides, by taking the limit $\eta\rightarrow1$ and $\chi=0$, we recover the result in Minkowski spacetime given in (\ref{2.19}). 

Finally, following the steps from Eq. (\ref{2.21}) to Eq. (\ref{2.24}), we can obtain the components of the Dirac spinor that are parallel and anti-parallel to the $z$ axis of the cosmic dislocation spacetime. Thus, the parallel component to the $z$ axis of the Dirac spinor for positive-energy solutions is
\begin{eqnarray}
\Psi_{+}&=&h_{+}\,\,_{1}F_{1}\left(-n,\frac{\left|\bar{\zeta}_{+}\right|}{\eta}+1;\frac{\mu\lambda}{2}\rho^{2}\right)\left(
\begin{array}{c}
1	\\
0 \\
\frac{k}{\left(\mathcal{E}+m\right)} \\
\frac{i\,e^{i\varphi}}{\left(\mathcal{E}+m\right)}\left(\mu\lambda\rho-\frac{\left|\bar{\zeta}_{+}\right|}{\eta\rho}+\frac{\bar{\zeta}_{+}}{\eta\rho}\right)\\
\end{array}\right)\nonumber\\
[-2mm]\label{5.12b}\\[-2mm]
&+&h_{+}\,\frac{i\,e^{i\varphi}}{\left(\mathcal{E}+m\right)}\frac{n\,\mu\lambda\rho}{\left(\frac{\left|\bar{\zeta}_{+}\right|}{\eta}+1\right)}\,_{1}F_{1}\left(-n+1,\frac{\left|\bar{\zeta}_{+}\right|}{\eta}+2;\frac{\mu\lambda}{2}\rho^{2}\right)\left(
\begin{array}{c}
0\\
0\\
0\\
1\\	
\end{array}\right),\nonumber
\end{eqnarray}
while the anti-parallel component to the $z$ axis of the cosmic dislocation of the Dirac spinor for positive-energy solutions is
\begin{eqnarray}
\Psi_{-}&=&h_{-}\,\,_{1}F_{1}\left(-n,\frac{\left|\bar{\zeta}_{-}\right|}{\eta}+1;\frac{\mu\lambda}{2}\rho^{2}\right)\left(
\begin{array}{c}
0	\\
1 \\
\frac{i\,e^{-i\varphi}}{\left(\mathcal{E}+m\right)}\left(\mu\lambda\rho-\frac{\left|\bar{\zeta}_{-}\right|}{\eta\rho}-\frac{\bar{\zeta}_{-}}{\eta\rho}\right)\\
-\frac{k}{\left(\mathcal{E}+m\right)} \\
\end{array}\right)\nonumber\\
[-2mm]\label{5.12c}\\[-2mm]
&+&h_{-}\,\frac{i\,e^{-i\varphi}}{\left(\mathcal{E}+m\right)}\frac{\,n\,\mu\lambda\rho}{\left(\frac{\left|\bar{\zeta}_{-}\right|}{\eta}+1\right)}\,_{1}F_{1}\left(-n+1,\frac{\left|\bar{\zeta}_{-}\right|}{\eta}+2;\frac{\mu\lambda}{2}\rho^{2}\right)\left(
\begin{array}{c}
0\\
0\\
1\\
0\\	
\end{array}\right),\nonumber
\end{eqnarray}
where we have defined the parameter $h_{\pm}=h_{s}$ in (\ref{5.12b}) and (\ref{5.12c}) as
\begin{eqnarray}
h_{s}=C\,e^{-i\mathcal{E}t}\,e^{i\left(l+\frac{1}{2}-\frac{s}{2}\right)\varphi}\,e^{ikz}\left(\frac{\mu\lambda}{2}\right)^{\frac{\left|\bar{\zeta}_{s}\right|}{2\eta}}\,e^{-\frac{\mu\lambda}{4}\rho^{2}}\,\rho^{\left|\bar{\zeta}_{s}\right|},
\label{5.12d}
\end{eqnarray} 
where $\bar{\zeta}_{s}=l+\frac{s}{2}\left(1-\eta\right)+\frac{1}{2}\left(1-s\right)-\chi k+s\frac{\phi_{\mathrm{AC}}}{2\pi}$. As discussed in the previous sections, Negative-energy solutions of the Dirac equation (\ref{4.8}) can also be obtained by applying the same procedure used to obtain Eqs. (\ref{5.12b}) and (\ref{5.12c}).

\section{Applications and Discussion}

In this section, we discuss some applications of the results obtained in the previous sections. We have obtained a spectrum of energy of the Dirac oscillator in the presence of Aharonov-Casher coupling in three distinct spacetime background. We observe that the potential confines the particle and the spectrum exhibit a dependence in the Aharonov-Casher geometric phase similar to that obtained for the energy spectrum of quantum dots in the presence of a Aharonov-Bohm flux \cite{ab}. In quantum dots, for instance quantum dots described by a parabolic confining potential \cite{tan}, a physical quantity arises from the presence of a Aharonov-Bohm flux in this system: the persistent current. Persistent currents are obtained by using the Byers-Yang relation \cite{by}. In the present case, we can obtain a similar persistent current by considering the Dirac oscillator playing the role of the confining potential analogous to a parabolic confining potential which describes a quantum dot potential. In the following, in order to obtain persistent currents that arise from the dependence of the Aharonov-Casher geometric phase studied previously, we write the Byers-Yang relation \cite{by,ring1} in the form:
\begin{eqnarray}
\mathcal{I}=-\sum_{n,\,l}\frac{\partial\mathcal{E}_{n,\,l}}{\partial\phi_{\mathrm{AC}}}.
\end{eqnarray}

Studies of the arising of persistent currents in quantum rings from the dependence of the energy levels on the Berry phase \cite{berry} and Aharonov-Anandan quantum phase \cite{ahan} have been made in Ref. \cite{ring2}. Moreover, persistent currents have been studied from dependence of the energy levels of the bound states on the Aharonov-Casher geometric phase in Refs. \cite{ring1,bf18}, which are called persistent spin currents. In the following, we calculate the persistent spin currents which arises from the dependence of the relativistic energy levels of the Dirac oscillator on the Aharonov-Casher geometric phase obtained in the previous sections.

\subsection{Minkowski spacetime case}

We have seen in (\ref{2.19}) that the relativistic energy levels of the Dirac oscillator depend on the Aharonov-Casher geometric phase \cite{ac}. Hence, from the dependence of the relativistic energy levels (\ref{2.19}) on the Aharonov-Casher geometric phase, we have the arising of persistent spin currents \cite{ring1,by} given by
\begin{eqnarray}
\mathcal{I}=-\sum_{n,\,l}\frac{\partial\mathcal{E}_{n,\,l}}{\partial\phi_{\mathrm{AC}}}=-\frac{m\omega}{2\pi}\,\sum_{n,\,l}\frac{\left(s\frac{\zeta_{s}}{\left|\zeta_{s}\right|}-1\right)}{\left[m^{2}+k^{2}+4m\omega\left(n+\frac{\left|\zeta_{s}\right|}{2}-\frac{\zeta_{s}}{2}\right)\right]^{-1/2}}.
\label{2.20}
\end{eqnarray}
Note that the expression of the persistent currents (\ref{2.20}) is a periodic function of the Aharonov-Casher geometric phase $\phi_{\mathrm{AC}}$.

\subsection{Cosmic string spacetime}

From the dependence of the energy levels (\ref{3.18}) on the Aharonov-Casher geometric phase $\phi_{\mathrm{AC}}$, we also have the arising of persistent spin currents \cite{by} given by
\begin{eqnarray}
\mathcal{I}=-\sum_{n,\,l}\frac{\partial\mathcal{E}_{n,\,l}}{\partial\phi_{\mathrm{AC}}}=-\frac{m\omega}{2\pi}\,\sum_{n,\,l}\frac{\left(s\frac{\tilde{\zeta}_{s}}{\left|\tilde{\zeta}_{s}\right|}-1\right)}{\left[m^{2}+k^{2}+4m\omega\left(n+\frac{\left|\tilde{\zeta}_{s}\right|}{2\eta}-\frac{\tilde{\zeta}_{s}}{2\eta}\right)\right]^{-1/2}}.
\label{3.19}
\end{eqnarray}

By comparing (\ref{3.19}) with the results of the previous section given in (\ref{2.20}), we have that the curvature of the conical surface changes the pattern of oscillations of the persistent currents which agrees with Ref. \cite{fur4}. Furthermore, by taking the limit $\eta\rightarrow1$, we can observe that we recover all results obtained in the Minkowski spacetime obtained in the previous section.

\subsection{Cosmic dislocation spacetime}

Again, we have that the dependence of the relativistic energy levels (\ref{4.9}) on the geometric quantum phase $\phi_{\mathrm{AC}}$ gives rise to the arising of persistent spin currents \cite{by}, which are given by
\begin{eqnarray}
\mathcal{I}=-\sum_{n,\,l}\frac{\partial\mathcal{E}_{n,\,l}}{\partial\phi_{\mathrm{AC}}}=-\frac{m\omega}{2\pi}\,\sum_{n,\,l}\frac{\left(s\frac{\bar{\zeta}_{s}}{\left|\bar{\zeta}_{s}\right|}-1\right)}{\left[m^{2}+k^{2}+4m\omega\left(n+\frac{\left|\bar{\zeta}_{s}\right|}{2\eta}-\frac{\bar{\zeta}_{s}}{2\eta}\right)\right]^{-1/2}}
\label{4.10}
\end{eqnarray}

We can observe in Eq. (\ref{4.10}) that the pattern of oscillations of the persistent currents changes in contrast to the previous results obtained in Eqs. (\ref{2.20}) and (\ref{3.19}). In this case, both torsion and curvature effects change the pattern of oscillations of the persistent spin currents. Note, by taking $\chi=0$, we recover the result obtained in Eq. (\ref{3.19}). Besides, by taking the limit $\eta\rightarrow1$ and $\chi=0$, we recover the result in Minkowski spacetime given in (\ref{2.20}).

\section{conclusions}

We have discussed the influence of the Aharonov-Casher effect \cite{ac} on the Dirac oscillator \cite{osc1} in three different scenarios of general relativity. By using the mathematical formulation of the spinor theory in curved space \cite{weinberg}, we have shown that we can solve the Dirac equation for the Dirac oscillator under the influence of the Aharonov-Casher effect. We have found the eigenvalues e eigenfunctions for the Dirac oscillator in three different scenarios: in the Minkowski spacetime, the cosmic string spacetime and the cosmic dislocation. As it was already noted in Ref. \cite{josevi}, the behaviour of the Dirac oscillator in the presence of the same topological defects considered here and the Aharonov-Bohm flux is characterized by energy levels that depend on the parameters related to the topological defects. Thus, we have shown that in all cases that the relativistic energy levels of bound states depend on the Aharonov-Casher geometric phase. We have also seen that both curvature and torsion break the degeneracy of the relativistic energy levels in the same way of the breaking of the degeneracy of the relativistic Landau levels given in Refs. \cite{furt7,bf11}. We have also studied an application of each system investigated here which consists in a relativistic quantum dot for neutral particles, where we consider the Aharonov-Casher coupling in the presence of a confining potential represented by Dirac oscillator. We have calculated the persistent currents in all cases investigated in this contribution and obtained that the persistent currents are periodic functions of the Aharonov-Casher geometric phase. In this way, we can suggest that this model is a relativistic generalization of the Aharonov-Casher quantum dot. It worth mentioning that the study of the Dirac oscillator in the presence of topological defects serve as a basis for future investigations of the Jaynes-Cummings model in the presence of a topological defect.

\acknowledgments{We would like to thank CNPq (Conselho Nacional de Desenvolvimento Cient\'ifico e Tecnol\'ogico - Brazil) for financial support.}


\end{document}